# A consistent variational formulation of Bishop nonlocal rods

Raffaele Barretta,

S. Ali Faghidian,

Francesco Marotti de Sciarra

# A consistent variational formulation of Bishop nonlocal rods

R. Barretta*, S. Ali Faghidian, F. Marotti de Sciarra

*Department of Structures for Engineering and Architecture, University of Naples Federico II, via Claudio 21, 80125 Naples, Italy – e-mails: rabarret@unina.it - faghidian@gmail.com - marotti@unina.it*

**Abstract**

Thick rods are employed in Nanotechnology to build modern electro mechanical systems. Design and optimization of such structures can be carried out by nonlocal continuum mechanics which is computationally convenient when compared to atomistic strategies. Bishop's kinematics is able to describe small-scale thick rods if a proper mathematical model of nonlocal elasticity is formulated to capture size effects. In all papers on the matter, nonlocal contributions are evaluated by replacing Eringen's integral convolution with the consequent (but not equivalent) differential equation governed by Helmholtz's differential operator. As notorious in integral equation theory, this replacement is possible for convolutions, defined in unbounded domains, governed by averaging kernels which are Green's functions of differential operators. Indeed, Eringen himself, in order to study nonlocal problems defined in unbounded domains, such as screw dislocations and wave propagations, suggested to replace integro-differential equations with differential conditions. A different scenario appears in Bishop rod mechanics where nonlocal integral convolutions are defined in bounded structural domains, so that Eringen's nonlocal differential equation has to be supplemented with additional boundary conditions. The objective is achieved by formulating the governing nonlocal equations by a proper variational statement. The new methodology provides an amendment of previous contributions in literature and is illustrated by investigating the elastostatic behavior of simple structural schemes. Exact solutions of Bishop rods are evaluated in terms of nonlocal parameter and cross-section gyration radius.

*Corresponding author
E-mail: rabarret@unina.it

Both hardening and softening structural responses are predictable with a suitable tuning of the parameters.

**1. Introduction**

Advances in nano-engineering have required the development of tools suitable for design and optimization of Nano-Electro-Mechanical Systems (NEMS). However, characteristic length-scale effects in nano-structures cannot be appropriately modeled by classical local methodologies [1]. Therefore, the thematic dealing with size-dependent mechanical analysis of scale phenomena in nano-structured components of NEMS is in the major focus of interest in the current literature [2-15], see also the review papers [16, 17].

Eringen in his paper [18] presented solutions of screw dislocation and surface waves in the framework of nonlocal integral theory of elasticity, where nonlocal stress fields were defined by integral convolutions of local stress fields with suitable averaging kernels. Nonlocal problems were formulated in [18] by selecting averaging kernels to be Green functions of differential operators. Consequently, the governing integro-differential nonlocal equations of elastic equilibrium were reverted to seemingly more convenient differential equations. Structural problems of technical interest are instead defined in bounded domains and thus Eringen integro-differential equations cannot be replaced with nonlocal differential equations, since appropriate constitutive boundary conditions (CBC) have to be prescribed to ensure equivalence [19]. Notwithstanding this and for almost twenty years, as apparent from the review paper [20], CBCs have been overlooked. Nonlocal structural problems have been addressed by improperly replacing Eringen's nonlocal integral model (EIM), equipped with Helmholtz kernel ([18], Eq. (3.17)), with the non-equivalent differential law [18], Eq. (3.19). Such an inaccurate choice has been source of paradoxes [21-24] and outcomes of limited applicative interest in structural mechanics [21-24]. These issues have been addressed in [25] for inflected beams, by showing that proper CBCs have to be imposed to close the relevant integral problem. Well-posed nonlocal integral elasticity models have been formulated by generalized functions in [26] and numerically solved by iterative techniques in [2]. Nowadays, it is acknowledged by the scientific community that Eringen's integral theory is inapplicable to structural problems governed by Bernoulli-Euler, Timoshenko beams and Kirchhoff plates due to conflict between equilibrium and nonlocal boundary conditions.

In the present paper, it is shown that, as an exception, Eringen's integral model does not clash against equilibrium if exploited to predict size effects in thick rods formulated by Bishop's kinematics [27]. Motivations the present research and mathematical inaccurateness of previous contributions on Bishop nonlocal rods are elucidated in the next section.

## 2. Motivation and outline

It is well-established that to adequately model thick rods, independently of structural scale, effects of lateral deformation (Rayleigh–Love theory) and shear stiffness (Bishop theory) are important and cannot be overlooked [28]. As underlined below, contributions in literature (see e.g. [29, 30]) on Bishop elastically nonlocal rods are based on non-equivalent differential constitutive laws associated with the original strain-driven Eringen integral convolution [18]. The differential law of Eringen nonlocal elasticity has been however acknowledged in Engineering Science to be inapt to evaluate size effects in nanostructures, see e.g. [22, 24].

The motivation of the present study is therefore to re-formulate the elastic equilibrium problem of a thick rod in an appropriate variational setting which provides proper constitutive boundary conditions, overlooked in previous contributions [29-41]. In papers dealing with nonlocal thick rods, EIM is preliminarily introduced as (see e.g. [30], Eq. (1a); [41], Eq. (1))

$$\boldsymbol{\sigma}(\mathbf{x}) = \int_V \varphi_\lambda\left(|\mathbf{x}-\mathbf{x}'|\right) \mathbf{C} : \boldsymbol{\varepsilon}(\mathbf{x}) dv \tag{1}$$

with $\boldsymbol{\sigma}$, $\boldsymbol{\varepsilon}$ and $\mathbf{C}$ denoting, respectively, nonlocal stress, local strain field and elastic stiffness. The averaging kernel $\varphi_\lambda$ incorporates nonlocality at a reference point $\mathbf{x}$ by the local stress at a source point $\mathbf{x}'$. Nevertheless, EIM has not been properly resorted to, being been incongruously substituted with the non-equivalent Eringen differential law ([30], Eq. (2); [41], Eq. (3))

$$\boldsymbol{\sigma}(\mathbf{x}) - L_c^2 \Delta\boldsymbol{\sigma}(\mathbf{x}) = \mathbf{C} : \boldsymbol{\varepsilon}(\mathbf{x}) \tag{2}$$

with $\Delta$ Laplace operator and $L_c$ nonlocal characteristic length. Equivalence between Eqs. (1) and (2) holds true for problems defined in unbounded domains, due to tacit fulfillment of vanishing involved fields at infinity [19]. For bounded structural domains of nano-mechanical interest, suitable constitutive boundary conditions (undetected in Bishop rod nonlocal theory) have to be instead detected and prescribed to ensure equivalence between Eqs. (1) and (2). Despite the numerous contributions in literature, the challenging issue of adopting a consistent nonlocal (integral) constitutive theory to significantly capture size effects in thick rods has not been appropriately addressed. In the present study, scale effects in Bishop rods are modelled by a novel variationally consistent approach. The integral convolution, formulated for Bishop rods, with an elastic strain field is demonstrated to generate a linear space of nonlocal stress fields having non-empty intersection with the affine set of equilibrated stress fields, and therefore leads to well-posed elastostatic applicative problems. The nonlocal Bishop rod model is established here with the intent of amending previous contributions on the matter. The new theory, exempt from drawbacks peculiar of strain-driven beam and plate nonlocal models, is capable to predict both hardening and softening structural responses.

The plan is the following. The local elasticity problem of Bishop rods is preliminarily reformulated in Sect. 3. The nonlocal integral elasticity model of Bishop rods is then developed in Sect. 4 by convoluting the local elasticity response functions provided in Sect. 3 with averaging kernels adopted by Eringen himself in [18]. An equivalent differential nonlocal problem, equipped with non-classical constitutive boundary conditions, is also provided. The methodology is presented in Sect. 5 by analytically investigating simple nonlocal Bishop rods subject to selected loading systems and kinematic boundary conditions. Benchmark solutions of nonlocal thick nanorods are detected, numerically analyzed and graphically outlined. Closing remarks are collected in Sect. 6.

## 3. Bishop rod in local elasticity

An elastically homogeneous thick rod of length $L$ with cross-section $\Omega$, subjected to a distributed axial force per unit length $p$ in the interval $[0,L]$, concentrated forces $\bar{N}$ at the rod ends are considered here. The abscissa along the rod axis is denoted by $x$ and the radial position of a cross-sectional point with respect to the centroid is denoted by $r$. As a result of studying an axisymmetric problem, the circumferential displacement $u_\theta$ is vanishing. According to the fundamental hypotheses of the Love rod model, radial stress may be assumed to vanish on the cross-section [42], and consequently, the displacement field, including lateral deformation effects, can be expressed as

$$u_x(x,r)=u(x), \qquad u_r(x,r)=-\nu r \partial_x u(x), \qquad u_\theta(x,r)=0 \qquad (3)$$

where $u_x$, $u_r$ and $u_\theta$ respectively denote the displacement components along $x-$, $r-$ and $\theta-$coordinate directions, along with $\nu$ being the Poisson ratio. The non-vanishing components of linearized strains according to the Love rod model are given by

$$\varepsilon_x = \partial_x u, \qquad \varepsilon_r = \varepsilon_\theta = -\nu \partial_x u, \qquad \gamma_{xr} = -\nu r \partial_x^2 u \qquad (4)$$

The Bishop rod model of local elasticity is governed by the following elastic energy

$$R^{Loc}(\varepsilon_x, \gamma_{xr}) := \frac{1}{2}\int_0^L \left(\int_\Omega \left(E\varepsilon_x^2 + G\gamma_{xr}^2\right)dA\right)dx \qquad (5)$$

which, taking Eq. (4) into account, can be expressed in terms of axial strain field $\varepsilon \in C^2([0,L];\Re)$ by

$$\begin{aligned} R^{Loc}(\varepsilon) &= \frac{1}{2}\int_0^L \left(\int_\Omega \left(E\varepsilon^2 + \frac{E}{2(1+\nu)}\nu^2 r^2 (\partial_x \varepsilon)^2\right)dA\right)dx \\ &= \frac{1}{2}\int_0^L EA\left(\varepsilon^2 + \frac{\nu^2 \rho^2}{2(1+\nu)}(\partial_x \varepsilon)^2\right)dx \end{aligned} \qquad (6)$$

The gyration radius $\rho$ is defined by

$$A\rho^2 = J := \int_\Omega r^2 dA \qquad (7)$$

being $J$ the polar moment of inertia about the cross-sectional geometric center. Shear and Euler-Young moduli are also denoted by $G$ and $E$, with $G = E/2(1+\nu)$. The corresponding stress field in a local thick rod is the axial force field $N \in C^1([0,L];\Re)$ which is obtained by the variational constitutive condition

$$\langle N, \delta\varepsilon \rangle := \int_0^L N(x)\delta\varepsilon(x)dx = \langle dR^{Loc}(\varepsilon), \delta\varepsilon \rangle \qquad (8)$$

for any virtual axial strain field $\delta\varepsilon \in C^1([0,L];\Re)$. Taking into account the expression of $R^{Loc}$ in Eq. (6), a direct evaluation provides the derivative of the elastic energy along a virtual axial strain field

$$\langle dR^{Loc}(\varepsilon), \delta\varepsilon \rangle = \int_0^L EA\,\varepsilon\delta\varepsilon\,dx + \int_0^L EA\,\frac{v^2\rho^2}{2(1+v)}(\partial_x\varepsilon)\partial_x(\delta\varepsilon)dx \qquad (9)$$

Integrating by parts Eq. (9) and imposing the variational condition Eq. (8), a standard localization procedure provides the following differential problem equipped with constitutive boundary conditions

$$\begin{cases} EA\left(\varepsilon - \dfrac{v^2\rho^2}{2(1+v)}\partial_x^2\varepsilon\right) = N, & \text{in } [0,L] \\ v^2\rho^2\partial_x\varepsilon = 0, & \text{on } \partial[0,L] \end{cases} \qquad (10)$$

where $\partial[0,L]$ is the boundary of the interval $[0,L]$.

Unlike contributions in literature, the local elastic model of thick rods is properly defined in terms of the axial force field which is appropriately related to the axial strain field and to its first-order gradient Eq. (10)$_1$. The developed local constitutive law of Bishop rod is equipped with univocally-detected constitutive boundary conditions Eq. (10)$_2$ which do not clash against the requirements dictated by equilibrium. It is worth remarking that the classical local elastic law of slender rods is obtained by assuming a vanishing Poisson ratio $v$ in Eq. (10). We recall the differential and boundary conditions of equilibrium of a rod are given by

$$\begin{aligned}&\partial_x N + p = 0 \\ &(N+\bar{N})\delta u\big|_{x=0} = (N-\bar{N})\delta u\big|_{x=L} = 0\end{aligned} \qquad (11)$$

As it can be noticeably inferred from the established boundary-value problem, the Bishop thick rod model comprises not only the lateral deformation but also the shear stiffness effects.

## 4. Bishop rod in nonlocal integral elasticity

Let us preliminarily recall the definition of integral convolution of a scalar field $s$ with a averaging kernel $\varphi_\lambda$

$$(\varphi_\lambda * s)_x = (\varphi_\lambda * s)(x) := \int_0^L \varphi_\lambda(x - x') s(x') dx' \qquad (12)$$

with $x$ and $x'$ points of the interval $[0, L]$ and the scalar kernel $\varphi_\lambda$ fulfilling positivity, parity, symmetry, normalization and limit impulsivity properties [25, 26].

Following the seminal treatment by Eringen [18], the nonlocal integral model of a Bishop rod is formulated by properly including convolutions in Eq. (6), generating thus the elastic energy

$$R^{NLoc}(\varepsilon) := \frac{1}{2} \int_0^L EA \left( (\varphi_\lambda * \varepsilon)_x \varepsilon + \frac{v^2 \rho^2}{2(1+v)} (\varphi_\lambda * \partial_x \varepsilon)_x \partial_x \varepsilon \right) dx \qquad (13)$$

It is worth remarking that the local elastic energy $R^{Loc}$ Eq. (6) is recovered as the nonlocal parameter $\lambda$ in Eq. (13) tends to zero.

The corresponding nonlocal axial force $N$ in a thick rod is obtained in variational terms by

$$\langle N, \delta\varepsilon \rangle := \int_0^L N(x) \delta\varepsilon(x) dx = \langle dR^{NLoc}(\varepsilon), \delta\varepsilon \rangle \qquad (14)$$

for any virtual axial strain field $\delta\varepsilon \in C_0^1([0, L]; \Re)$ having compact support in $[0, L]$.

Taking into account the expression of $R^{NLoc}$ in Eq. (13) and integrating by parts, the directional derivative of the elastic energy is provided by

$$\left\langle dR^{NLoc}(\varepsilon), \delta\varepsilon \right\rangle = EA\left(\int_0^L (\varphi_\lambda * \varepsilon)_x \, \delta\varepsilon\, dx + \frac{v^2 \rho^2}{2(1+v)} \int_0^L (\varphi_\lambda * \partial_x \varepsilon)_x \, \partial_x(\delta\varepsilon)\, dx \right)$$

$$= EA\left(\int_0^L (\varphi_\lambda * \varepsilon)_x \, \delta\varepsilon\, dx - \frac{v^2 \rho^2}{2(1+v)} \int_0^L \partial_x (\varphi_\lambda * \partial_x \varepsilon)_x \, \delta\varepsilon\, dx \right. \quad (15)$$

$$\left. + \frac{v^2 \rho^2}{2(1+v)} \left( (\varphi_\lambda * \partial_x \varepsilon)_L \, \delta\varepsilon_L - (\varphi_\lambda * \partial_x \varepsilon)_0 \, \delta\varepsilon_0 \right) \right)$$

Since the test fields $\delta\varepsilon \in C_0^1([0,L]; \Re)$ in the nonlocal variational equation Eq. (14) have compact supports, the boundary values $\delta\varepsilon_L := \delta\varepsilon(L)$ and $\delta\varepsilon_0 := \delta\varepsilon(0)$ are vanishing. Accordingly, the boundary terms in Eq. (15) disappear.

Enforcing the variational condition Eq. (14), a standard localization procedure provides the expression of the nonlocal axial force field $N$ in terms of the elastic axial strain field $\varepsilon_x$

$$N(x) = EA\left( (\varphi_\lambda * \varepsilon)(x) - \frac{v^2 \rho^2}{2(1+v)} \partial_x (\varphi_\lambda * \partial_x \varepsilon)(x) \right)$$

$$= EA\left( \int_0^L \varphi_\lambda(x - x')\varepsilon(x')\,dx' - \frac{v^2 \rho^2}{2(1+v)} \partial_x \left( \int_0^L \varphi_\lambda(x - x') \partial_{x'} \varepsilon(x')\,dx' \right) \right) \quad (16)$$

The nonlocal theory of elasticity for thick rods is therefore established by making recourse to a consistent variational constitutive formulation equipped with suitably chosen test fields.

It is worth noting that Eq. (16) has formally the same mathematical structure of the well-posed modified nonlocal strain gradient integral model for Bernoulli-Euler elastic beams (Barretta and Marotti de Sciarra [10], Eq. (1)). Accordingly, equipping the nonlocal integral convolution Eq. (16) with the special bi-exponential kernel, depicted in Fig. 1

$$\varphi_\lambda(x) := \frac{1}{2L_c} \exp\left( -\frac{|x|}{L_c} \right), \quad (17)$$

with characteristic length $L_c$ expressing the amplitude of the range of nonlocal action, we get the following significant result of constitutive equivalence.

**Proposition 4.1.** *The nonlocal integral law Eq. (16), with the bi-exponential kernel Eq. (17), is equivalent to the differential relation*

$$EA\left(\varepsilon(x) - \frac{v^2\rho^2}{2(1+v)}\partial_x^2\varepsilon(x)\right) = N(x) - L_c^2\partial_x^2 N(x) \qquad (18)$$

equipped with the constitutive boundary conditions (CBC)

$$\begin{cases}\partial_x N(0) = \dfrac{1}{L_c}N(0) + EA\dfrac{v^2\rho^2}{2(1+v)}\dfrac{1}{L_c^2}\partial_x\varepsilon(0) \\ \partial_x N(L) = -\dfrac{1}{L_c}N(L) + EA\dfrac{v^2\rho^2}{2(1+v)}\dfrac{1}{L_c^2}\partial_x\varepsilon(L)\end{cases} \qquad (19)$$

For bounded domains of applicative interest in nano-mechanics, constitutive boundary conditions Eq. (19) must be imposed to ensure the closure of the constitutive law Eq. (18), and therefore, assurance of the equivalence to the integral convolution law Eq. (16). The profound difference between the presented nonlocal formulation and the counterpart formulation of literature essentially lies on appropriately introducing and prescribing the CBCs in consistent mathematical form.

CBCs in Eq. (19) will be also demonstrated in Sect. 5 not to conflict with the equilibrium equations of rods Eq. (11). Thus, nonlocal axial force field output by the nonlocal convolutions can effectively meet the requirement of fulfilling conditions of equilibrium. Accordingly, the elastostatic problem of thick rods associated with the nonlocal integral elasticity model is well-posed. The local formulation of elastic thick rods as Eq. (10) can be recovered as the nonlocal characteristic length approaches zero in Eqs. (18)-(19).

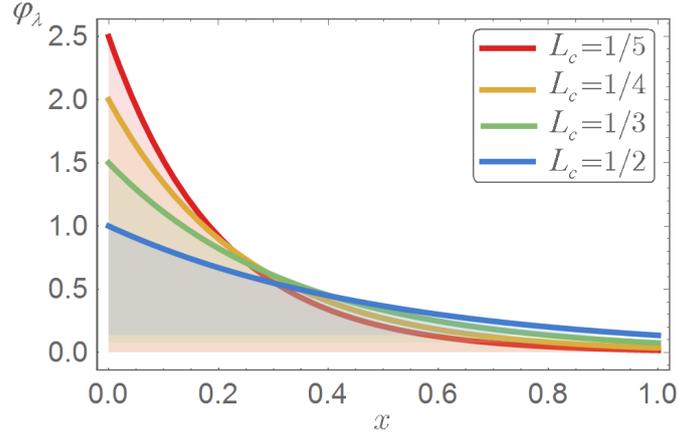

**Fig. 1.** Special bi-exponential kernel Eq. (17).

## 5. Case-studies

The nonlocal integral model is adopted here to examine the elastostatic response of a Bishop thick nano-rod subjected to a variety of loading systems and kinematic boundary conditions. For nano-rods under uniform axial load with intensity $\bar{p}$, the non-dimensional parameters: axial abscissa $\bar{x}$, characteristic nonlocal parameter $\lambda$, axial displacement $\bar{u}$ and radius of gyration $\bar{\rho}$, are introduced as

$$\bar{x} = \frac{x}{L}, \qquad \lambda = \frac{L_c}{L}, \qquad \bar{u}(\bar{x}) = u(x)\frac{EA}{\bar{p}L^2}, \qquad \bar{\rho}^2 = \frac{v^2 \rho^2}{2L^2(1+v)} \tag{20}$$

The non-dimensional displacement $\bar{u}$ for a rod subject to an axial tip-load $\bar{N}$ is defined by

$$\bar{u}(\bar{x}) = u(x)\frac{EA}{\bar{N}L} \tag{21}$$

To detect the elastostatic response of Bishop nonlocal nano-rods, the differential condition of equilibrium Eq. (11)$_1$ is firstly integrated, and consequently, the axial force field is determined in terms of an integration constant $\Upsilon_1$ as

$$N(x) = -\int_0^x p(\zeta)d\zeta + \Upsilon_1 \tag{22}$$

The constitutive differential equation of EDM Eq. (18) is subsequently solved to express the axial strain field $\varepsilon_x$ in terms of integration constants $\Upsilon_2, \Upsilon_3$ as

$$\varepsilon(x) = \Upsilon_2 \exp\left(\frac{x}{\bar{\rho}}\right) + \Upsilon_3 \exp\left(-\frac{x}{\bar{\rho}}\right) - \frac{1}{2EAL\bar{\rho}} \exp\left(\frac{x}{\bar{\rho}}\right) \int_0^x \exp\left(-\frac{\xi}{L\bar{\rho}}\right) \left(N(\xi) - L_c^2 \partial_\xi^2 N(\xi)\right) d\xi \\ + \frac{1}{2EAL\bar{\rho}} \exp\left(-\frac{x}{\bar{\rho}}\right) \int_0^x \exp\left(\frac{\eta}{L\bar{\rho}}\right) \left(N(\eta) - L_c^2 \partial_\eta^2 N(\eta)\right) d\eta \quad (23)$$

The axial displacement field $u$ is determined by integrating the differential condition of kinematic compatibility Eq. (4)₁ in terms of the integration constant $\Upsilon_4$ as

$$u(x) = \int_0^x \varepsilon(\zeta) d\zeta + \Upsilon_4 \quad (24)$$

The integration constants $\Upsilon_k (k=1..4)$ can be obtained by prescribing the standard kinematic and static boundary conditions (BC), along with the CBC Eq. (19). The exposed solution strategy provides exact elastic solutions by integrating differential equations of lower order.

Three different case studies will be examined here including uniformly loaded nano-rods with fixed-fixed and fixed-free ends as well as fixed-free nano-rods subjected to a tensile concentrated axial force at the free end.

### 5.1. *Uniformly loaded nano-rod with fixed-fixed ends*

For a nano-rod subject to a uniform axial loading with fixed-fixed ends, the kinematic boundary conditions (BC) are given by

$$u(0) = 0, \qquad u(L) = 0 \quad (25)$$

As a result of employing the solution technique described above, while prescribing standard BC and higher-order CBC, the non-dimensional axial displacement of nano-rod is determined

$$\bar{u}(\bar{x}) = \frac{1}{2}\left(\lambda + 2\lambda^2 + \bar{x} - \bar{x}^2 - 2\bar{\rho}^2 - \exp\left(\frac{\bar{x}}{\bar{\rho}}\right)\frac{\lambda + 2\lambda^2 - 2\bar{\rho}^2}{1+\exp\left(\frac{1}{\bar{\rho}}\right)} - \exp\left(\frac{1-\bar{x}}{\bar{\rho}}\right)\frac{\lambda + 2\lambda^2 - 2\bar{\rho}^2}{1+\exp\left(\frac{1}{\bar{\rho}}\right)}\right) \quad (26)$$

The non-dimensional axial displacement field according to the local Bishop rod model (LBM) can be also detected by setting $\lambda \to 0^+$ as

$$\bar{u}_{LBM}(\bar{x}) = \frac{1}{2}\left(\bar{x} - \bar{x}^2 - 2\bar{\rho}^2 + 2\bar{\rho}^2\cosh\left(\frac{1-2\bar{x}}{2\bar{\rho}}\right)\text{sech}\left(\frac{1}{2\bar{\rho}}\right)\right) \quad (27)$$

For numerical illustrations, the maximum value of axial displacement field is obtained

$$\bar{u}_{max} = \bar{u}\left(\bar{x} = \frac{1}{2}\right) = \frac{1}{8}\left(1 + 4\lambda + 8\lambda^2 - 8\bar{\rho}^2 - 4\left(\lambda + 2\lambda^2 - 2\bar{\rho}^2\right)\text{sech}\left(\frac{1}{2\bar{\rho}}\right)\right) \quad (28)$$

### 5.2. Uniformly loaded nano-rod with fixed-free ends

In case of a uniformly loaded nano-rod with fixed-free ends, the classical BC are expressed as

$$u(0) = 0, \qquad N(L) = 0 \quad (29)$$

The non-dimensional axial displacement of the nano-rod can be then detected exploiting the aforementioned solution technique while imposing the classical BC and CBC as

$$\bar{u}(\bar{x}) = \lambda + \lambda^2 + \bar{x} - \frac{\bar{x}^2}{2} - \bar{\rho}^2 - \left(\lambda + \lambda^2 - \bar{\rho}^2\right)\cosh\left(\frac{\bar{x}}{\bar{\rho}}\right) \\ + \left(\bar{\rho}^2 - \lambda^2 + \left(\lambda + \lambda^2 - \bar{\rho}^2\right)\cosh\left(\frac{1}{\bar{\rho}}\right)\right)\text{csch}\left(\frac{1}{\bar{\rho}}\right)\sinh\left(\frac{\bar{x}}{\bar{\rho}}\right) \quad (30)$$

The non-dimensional axial displacement consistent with LBM is also detected as $\lambda \to 0^+$

$$\bar{u}_{LBM}(\bar{x}) = \bar{x} - \frac{\bar{x}^2}{2} - \bar{\rho}^2 + \bar{\rho}^2\cosh\left(\frac{1-2\bar{x}}{2\bar{\rho}}\right)\text{sech}\left(\frac{1}{2\bar{\rho}}\right) \quad (31)$$

The maximum axial displacement is additionally determined for numerical presentation

$$\bar{u}_{\max} = \bar{u}(\bar{x}=1) = \frac{1}{2} + \lambda \tag{32}$$

The maximum axial deformation of the nonlocal Bishop nano-rod at the free end is independent of the non-dimensional radius of gyration $\bar{\rho}$. Accordingly, to examine the effects of non-dimensional gyration radius on the axial displacement filed, the value of the axial displacement field at the mid-span of the rod is considered

$$\bar{u}\left(\bar{x}=\frac{1}{2}\right) = \frac{1}{8}\left(3 + 8\lambda + 8\lambda^2 - 8\bar{\rho}^2 - 4(\lambda + 2\lambda^2 - 2\bar{\rho}^2)\operatorname{sech}\left(\frac{1}{2\bar{\rho}}\right)\right) \tag{33}$$

### 5.3. Tip-loaded nano-rod with fixed-free ends

When a nano-rod with fixed-free ends is subject to a tensile concentrated axial force $\bar{N}$ at the free end, the classical BC are given by

$$u(0) = 0, \qquad N(L) = \bar{N} \tag{34}$$

The non-dimensional axial displacement of the nano-rod is similarly evaluated utilizing the proposed solution technique, while imposing the classical BC and higher-order CBC as

$$\bar{u}(\bar{x}) = \lambda + \bar{x} - \lambda \operatorname{csch}\left(\frac{1}{2\bar{\rho}}\right)\sinh\left(\frac{1-2\bar{x}}{2\bar{\rho}}\right) \tag{35}$$

The non-dimensional local axial displacement field can be determined by letting $\lambda \to 0^+$,

$$\bar{u}_{\text{LBM}}(\bar{x}) = \bar{x} \tag{36}$$

For sake of numerical illustrations, the maximum axial displacement is obtained

$$\bar{u}_{\max} = \bar{u}(\bar{x}=1) = 1+2\lambda \tag{37}$$

which is independent of the non-dimensional gyration radius $\bar{\rho}$. Therefore, to study the effects of non-dimensional gyration radius on the axial displacement field, the value of the axial displacement field at $\bar{x}=1/4$ is considered

$$\bar{u}\left(\bar{x}=\frac{1}{4}\right) = \frac{1}{4} + \lambda - \frac{1}{2}\lambda\,\mathrm{sech}\left(\frac{1}{4\bar{\rho}}\right) \tag{38}$$

5.4. *Numerical illustrations*

3D plots of the non-dimensional axial displacement field $\bar{u}$ corresponding to the nonlocal integral elasticity model for uniformly loaded nano-rods with fixed-fixed and fixed-free ends and tip-loaded nano-rods with fixed-free ends are illustrated in Figs. 2-4. The assumed non-dimensional radius of gyration is $\bar{\rho}=0.2$ and the nonlocal parameter $\lambda$ is ranging in $]0,0.2[$. It is noticeably deduced from Figs. 2-4 that the adopted nonlocal integral model exhibits a softening behavior in terms of the characteristic parameter $\lambda$, increasing the axial displacement filed with increase of the parameter $\lambda$. The size-dependent elastostatic response of nonlocal Bishop nano-rods coincides with the results of local Bishop rod for vanishing characteristic parameter $\lambda \to 0^{+}$.

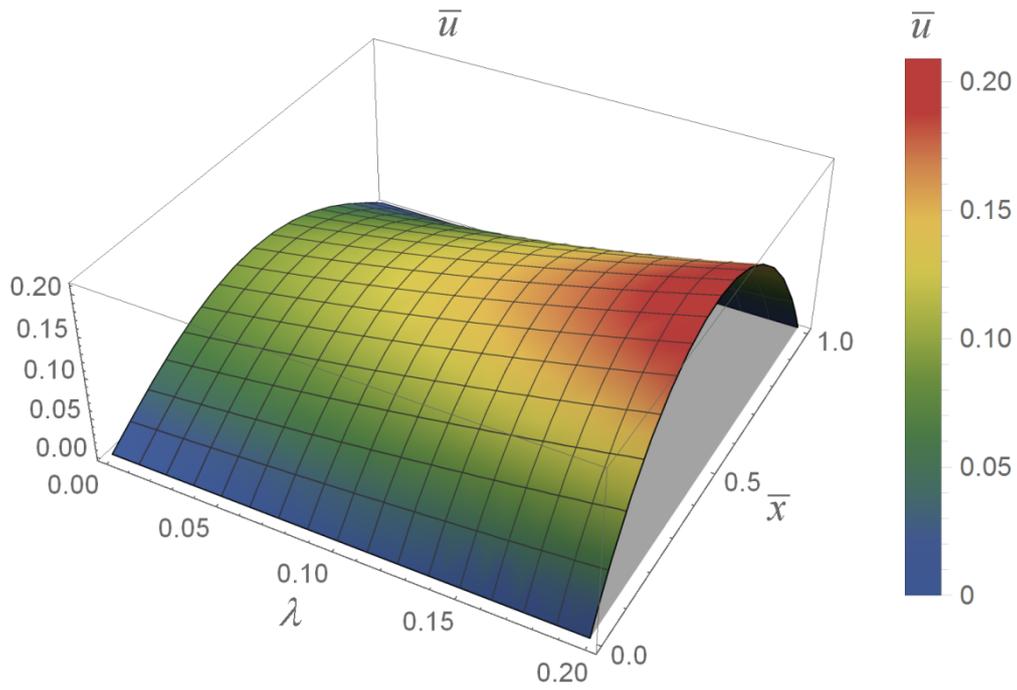

**Fig. 2.** Uniformly loaded nano-rod with fixed-fixed ends: $\bar{u}$ vs. $\bar{x}$ and $\lambda$ for $\bar{\rho} = 0.2$

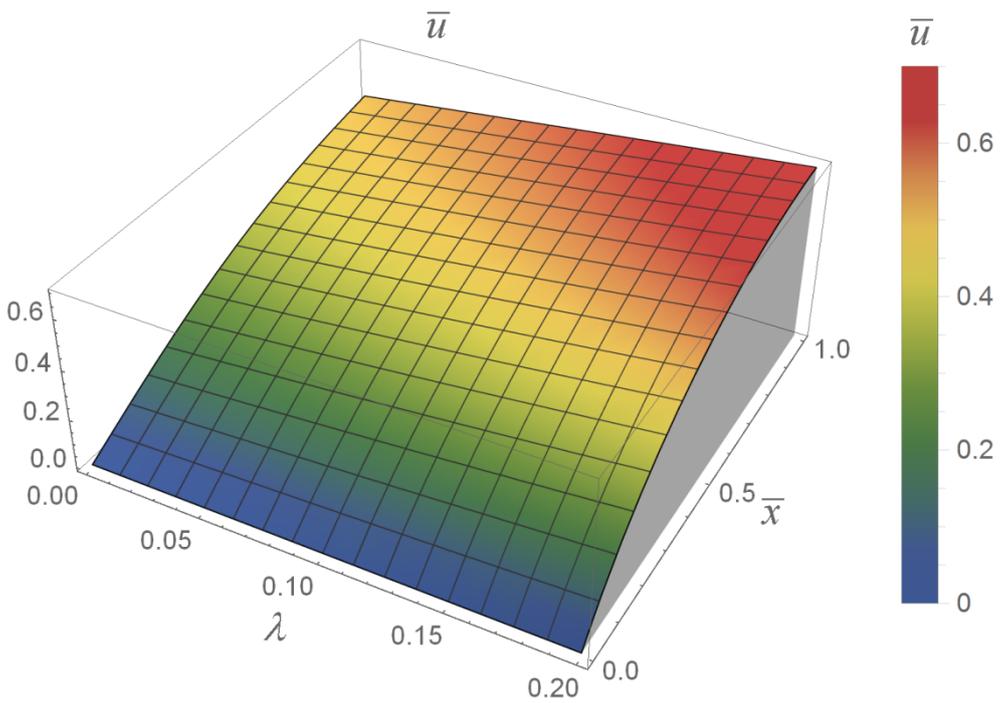

**Fig. 3.** Uniformly loaded nano-rod with fixed-free ends: $\bar{u}$ vs. $\bar{x}$ and $\lambda$ for $\bar{\rho} = 0.2$

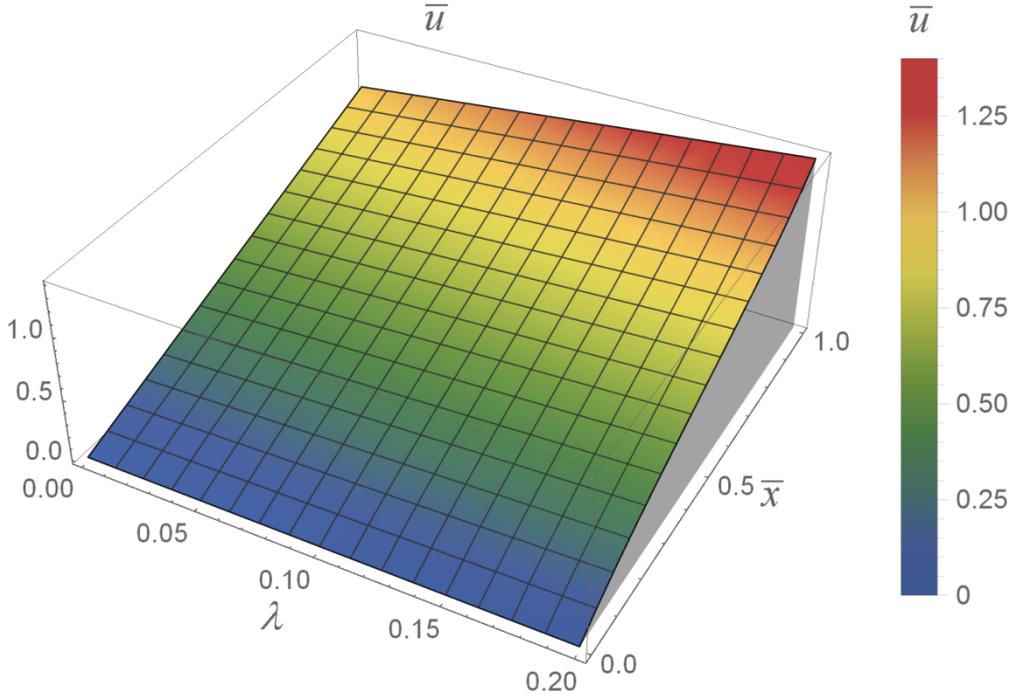

**Fig. 4.** Tip-loaded nano-rod with fixed-free ends: $\bar{u}$ vs. $\bar{x}$ and $\lambda$ for $\bar{\rho} = 0.2$

Figs. 5-7 depict the effects of the non-dimensional radius of gyration on the non-dimensional axial deformation $\bar{u}$ of uniformly loaded nano-rods with fixed-fixed and fixed-free ends as well as tip-loaded nano-rods with fixed-free ends. Two values for the nonlocal parameter $\lambda$ are assumed as $\lambda = 0.1, 0.2$ and the radius of gyration is ranging in $]0.1, 0.4[$. It is inferred from Figs. 5-7 that the axial displacement field reveals a stiffening behavior vs. the gyration radius. The maximum displacement in both uniformly loaded and tip-loaded fixed-free nano-rods is independent of the gyration radius. The numerical values of the maximum displacement $\bar{u}_{max}$ of uniformly loaded rods with fixed-fixed ends are reported in Table 1. For uniformly loaded nano-rods with fixed-free ends, numerical values of the maximum displacement $\bar{u}_{max}$ and the displacement at the rod mid-span are collected in Tables 2 and 3. In case of tip-loaded nano-rods with fixed-free ends, numerical values of the maximum axial deformation $\bar{u}_{max}$ and the axial displacement at $\bar{x} = 1/4$ are also listed in Tables 4 and 5.

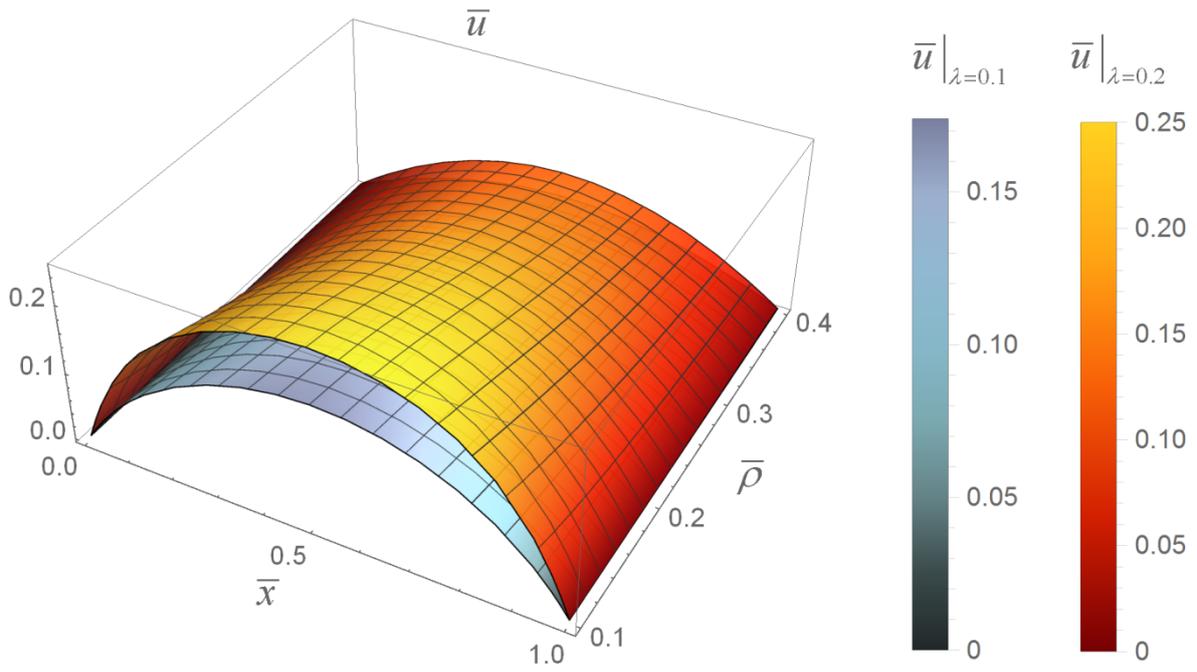

**Fig. 5.** Uniformly loaded nano-rod with fixed-fixed ends: $\bar{u}$ vs. $\bar{x}$ and $\bar{\rho}$ for $\lambda = 0.1, 0.2$

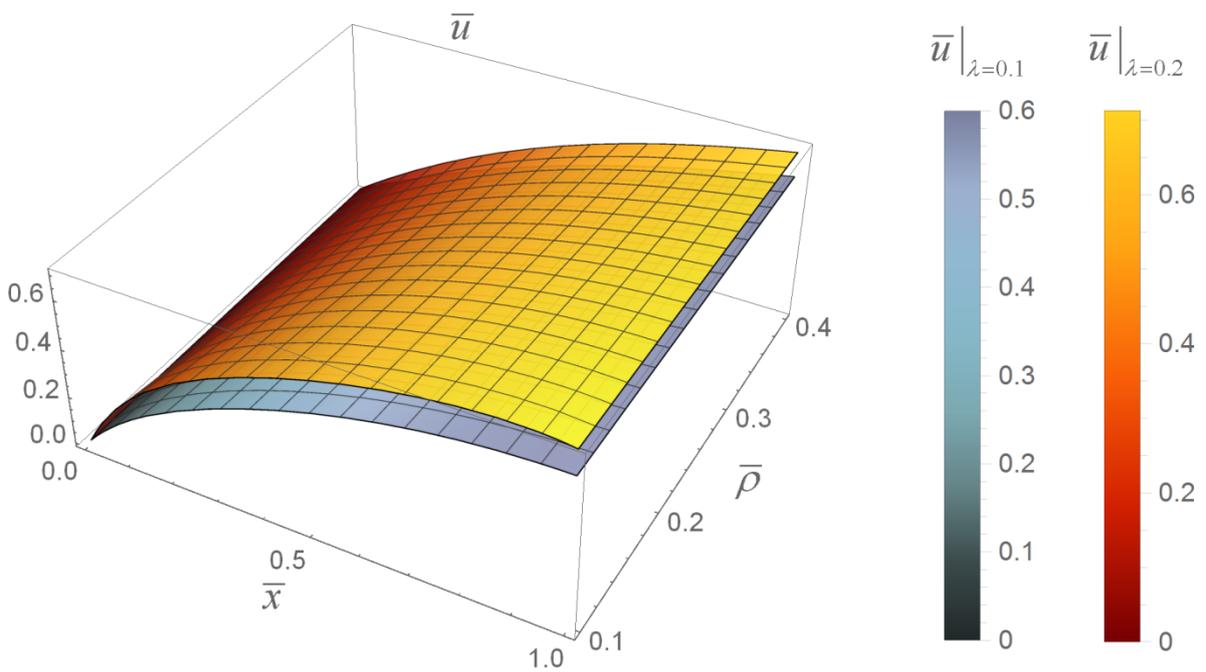

**Fig. 6.** Uniformly loaded nano-rod with fixed-free ends: $\bar{u}$ vs. $\bar{x}$ and $\bar{\rho}$ for $\lambda = 0.1, 0.2$

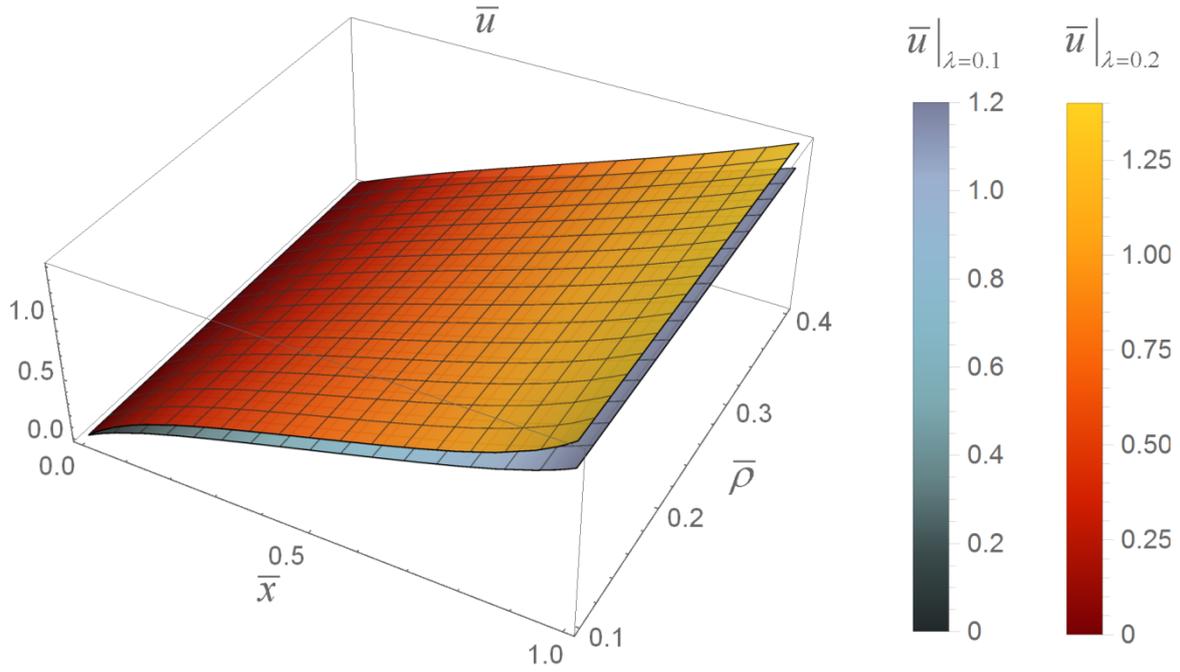

**Fig. 7.** Tip-loaded nano-rod with fixed-free ends: $\bar{u}$ vs. $\bar{x}$ and $\bar{\rho}$ for $\lambda = 0.1, 0.2$

## 6. Conclusion

The foremost outcomes of the present study can be enumerated as follows.

- The local elasticity model of Bishop rod has been recalled and improved in the present study to properly express the axial force field in terms of axial strain field and of its first-order gradient. The physically unmotivated definition of stress resultant (Eq. (10)$_2$ in [30]), as typically adopted in literature, has been thus shown to be unnecessary.

- Classical nonlocal formulations of Bishop elastic rods are founded on differential relations governed by Laplace's operator acting on stress and elastic strain fields, ignoring suitable non-classical constitutive boundary conditions. For bounded structural domains, such differential laws have been shown to be not equivalent to integral convolutions, since the constitutive boundary conditions are not automatically verified.

Eringen's nonlocal integral elasticity theory has been accurately applied to Bishop rods, amending thus previous papers which overlook the constitutive boundary conditions.

- The differential equations governing the nonlocal structural problem of a thick nano-rod have been established and equipped with the new constitutive boundary conditions which ensure equivalency between integral and differential formulations.

- The established nonlocal thick rod theory has been shown to be exempt from drawbacks of strain-driven nonlocal models applied to Bernoulli-Euler and Timoshenko beams.

- An effective solution procedure, based on integration of equations of lower order, has been applied to examine Eringen's nonlocal integral responses of simple thick nanorods. Exact nonlocal solutions have been also detected, numerically evaluated, graphically evidenced and commented upon.

- The proposed integral model has been shown to be able to predict both hardening and softening structural responses by suitably tuning cross-sectional gyration radius and small-scale parameter. Accordingly, the new approach could be usefully applied for design of new-generation nano-structured thick rod-like elements for NEMS problems.

**Acknowledgement**

The financial support of the Italian Ministry for University and Research (P.R.I.N. National Grant 2017, Project code 2017J4EAYB; University of Naples Federico II Research Unit) is gratefully acknowledged.

**Table 1.** Uniformly loaded Bishop nano-rod with fixed-fixed ends: $\bar{u}_{max}$ vs. $\lambda$

| | $\bar{u}_{max}$ | | | | | | | | | | |
|---|---|---|---|---|---|---|---|---|---|---|---|
| $\lambda$ | $0^+$ | 0.02 | 0.04 | 0.06 | 0.08 | 0.1 | 0.12 | 0.14 | 0.16 | 0.18 | 0.2 |
| $\bar{\rho} \to 0^+$ | 0.125 | 0.1354 | 0.1466 | 0.1586 | 0.1714 | 0.185 | 0.1994 | 0.2146 | 0.2306 | 0.2474 | 0.265 |
| $\bar{\rho} = 0.1$ | 0.115135 | 0.125395 | 0.136444 | 0.148282 | 0.160909 | 0.174326 | 0.188532 | 0.203527 | 0.219312 | 0.235885 | 0.253248 |
| $\bar{\rho} = 0.2$ | 0.0915233 | 0.100227 | 0.109601 | 0.119644 | 0.130356 | 0.141739 | 0.15379 | 0.166512 | 0.179903 | 0.193963 | 0.208693 |
| $\bar{\rho} = 0.3$ | 0.0678269 | 0.074433 | 0.081548 | 0.089171 | 0.097303 | 0.105942 | 0.11509 | 0.124746 | 0.13491 | 0.145582 | 0.156763 |
| $\bar{\rho} = 0.4$ | 0.049727 | 0.05462 | 0.059889 | 0.065534 | 0.071556 | 0.077954 | 0.084729 | 0.09188 | 0.099407 | 0.107311 | 0.115591 |

**Table 2.** Uniformly loaded Bishop nano-rod with fixed-free ends: $\bar{u}_{max}$ vs. $\lambda$

|  | $\bar{u}_{max}$ | | | | | | | | | |
|---|---|---|---|---|---|---|---|---|---|---|
| $\lambda$ | $0^+$ | 0.02 | 0.04 | 0.06 | 0.08 | 0.1 | 0.12 | 0.14 | 0.16 | 0.18 | 0.2 |
| For all $\bar{\rho}$ | 0.5 | 0.52 | 0.54 | 0.56 | 0.58 | 0.6 | 0.62 | 0.64 | 0.66 | 0.68 | 0.7 |

**Table 3.** Uniformly loaded Bishop nano-rod with fixed-free ends: $\bar{u}\left(\bar{x}=1/2\right)$ vs. $\lambda$

| | $\bar{u}\left(\bar{x}=1/2\right)$ | | | | | | | | | | |
|---|---|---|---|---|---|---|---|---|---|---|---|
| $\lambda$ | $0^+$ | 0.02 | 0.04 | 0.06 | 0.08 | 0.1 | 0.12 | 0.14 | 0.16 | 0.18 | 0.2 |
| $\bar{\rho} \to 0^+$ | 0.375 | 0.3954 | 0.4166 | 0.4386 | 0.4614 | 0.485 | 0.5094 | 0.5346 | 0.5606 | 0.5874 | 0.615 |
| $\bar{\rho} = 0.1$ | 0.365136 | 0.385395 | 0.406444 | 0.428282 | 0.450909 | 0.474326 | 0.498532 | 0.523527 | 0.549312 | 0.575885 | 0.603248 |
| $\bar{\rho} = 0.2$ | 0.341524 | 0.360227 | 0.379601 | 0.399644 | 0.420356 | 0.441739 | 0.46379 | 0.486512 | 0.509903 | 0.533963 | 0.558693 |
| $\bar{\rho} = 0.3$ | 0.317827 | 0.334433 | 0.351548 | 0.369171 | 0.387303 | 0.405942 | 0.42509 | 0.444746 | 0.46491 | 0.485582 | 0.506763 |
| $\bar{\rho} = 0.4$ | 0.299727 | 0.31462 | 0.329889 | 0.345534 | 0.361556 | 0.377954 | 0.394729 | 0.41188 | 0.429407 | 0.447311 | 0.465591 |

**Table 4.** Tip loaded Bishop nano-rod with fixed-free ends: $\bar{u}_{max}$ vs. $\lambda$

| $\lambda$ | $0^+$ | 0.02 | 0.04 | 0.06 | 0.08 | 0.1 | 0.12 | 0.14 | 0.16 | 0.18 | 0.2 |
|---|---|---|---|---|---|---|---|---|---|---|---|
| | | | | $\bar{u}_{max}$ | | | | | | | |
| For all $\bar{\rho}$ | 1 | 1.04 | 1.08 | 1.12 | 1.16 | 1.2 | 1.24 | 1.28 | 1.32 | 1.36 | 1.4 |

**Table 5.** Tip loaded Bishop nano-rod with fixed-free ends: $\bar{u}\left(\bar{x}=1/4\right)$ vs. $\lambda$

| | $\bar{u}\left(\bar{x}=1/4\right)$ | | | | | | | | | | |
|---|---|---|---|---|---|---|---|---|---|---|---|
| $\lambda$ | $0^+$ | 0.02 | 0.04 | 0.06 | 0.08 | 0.1 | 0.12 | 0.14 | 0.16 | 0.18 | 0.2 |
| $\bar{\rho} \to 0^+$ | 0.25 | 0.27 | 0.29 | 0.31 | 0.33 | 0.35 | 0.37 | 0.39 | 0.41 | 0.43 | 0.45 |
| $\bar{\rho} = 0.1$ | 0.25 | 0.268369 | 0.286739 | 0.305108 | 0.323477 | 0.341846 | 0.360216 | 0.378585 | 0.396954 | 0.415324 | 0.433693 |
| $\bar{\rho} = 0.2$ | 0.25 | 0.264705 | 0.279409 | 0.294114 | 0.308818 | 0.323523 | 0.338227 | 0.352932 | 0.367637 | 0.382341 | 0.397046 |
| $\bar{\rho} = 0.3$ | 0.25 | 0.262689 | 0.275378 | 0.288067 | 0.300756 | 0.313445 | 0.326134 | 0.338822 | 0.351511 | 0.3642 | 0.376889 |
| $\bar{\rho} = 0.4$ | 0.25 | 0.261679 | 0.273358 | 0.285036 | 0.296715 | 0.308394 | 0.320073 | 0.331752 | 0.343431 | 0.355109 | 0.366788 |